\documentclass[aps,prl,superscriptaddress,reprint,showpacs,floatfix]{revtex4-2}
\usepackage{graphicx}
\usepackage{dcolumn}
\usepackage{bm}
\usepackage{xcolor}
\usepackage{relsize}
\usepackage{amsmath}
\usepackage{amsfonts}
\usepackage{amssymb}
\usepackage{mathtools}
\usepackage{braket}
\usepackage{gensymb}
\usepackage{url}
\usepackage{mathrsfs}
\usepackage[colorlinks = true,
            linkcolor = blue,
            urlcolor  = blue,
            citecolor = blue,
            anchorcolor = blue]{hyperref}
\usepackage{footmisc}
\usepackage[utf8]{inputenc}
\usepackage{textcomp, gensymb}

\begin{document}

\preprint{APS/123-QED}

\title{Quantum hydrodynamics of a polariton fluid: pure energy relaxation terms}

\author{D. A. Saltykova}
\affiliation{ITMO University, St. Petersburg 197101, Russia}

\author{A. V. Yulin}
\affiliation{ITMO University, St. Petersburg 197101, Russia}

\author{I. A. Shelykh}

\affiliation{Science Institute, University of Iceland, Dunhagi 3, IS-107 Reykjavik, Iceland}

\date{\today}

\begin{abstract}

Cavity polaritons, hybrid half-light half-matter excitations in quantum microcavities in the strong-coupling regime demonstrate clear signatures of quantum collective behavior, such as analogues of Bose-Einstein condensate and superfluidity at remarkably high temperatures. The analysis of the formation of these states demands an account of the relaxation processes in the system. Although there are well-established approaches for the description of some of them, such as finite lifetime polariton, an external optical pump, and coupling with an incoherent excitonic reservoir, the treatment of pure energy relaxation in a polariton fluid still remains a puzzle. Here, based on the quantum hydrodynamics approach, we derive the corresponding equations where the energy relaxation term appears naturally. We analyze in detail how it affects the dynamics of polariton droplets and the dispersion of elementary excitations of a uniform polariton condensate. Although we focus on the case of cavity polaritons, our approach can be applied to other cases of bosonic condensates, where the processes of energy relaxation play an important role.
\end{abstract}

\maketitle

The physics of quantum fluids represents a major part of modern condensed matter and atomic physics. In the low-temperature limit the ensembles of identical quantum particles can form macroscopically coherent states corresponding to Bose-Einstein condensates (BECs) and demonstrate the remarkable property of superfluidity. This phenomenon is well studied in the domain of cold atomic gases, where the dynamics of condensate droplets can be modeled by the Gross-Pitaevskii equation for the macroscopic wavefunction (order parameter) of the condensate \cite{PitaevskiiBook}:
\begin{align}
    i\hbar\frac{\partial\psi}{\partial t} =-\frac{\hbar^2}{2m}\nabla^2\psi+U\psi+g|\psi|^2\psi.\label{GPEq}
\end{align}

Here, \(\psi(\mathbf{r}, t)\) is the macroscopic wavefunction of the condensate, \(\hbar\) is the reduced Planck constant, \(m\) is the mass of the particles, \(U(\mathbf{r})\) is the external trapping potential, and \(g\) is the interaction coefficient proportional to the \(s\)-wave scattering length. The first term on the right-hand side corresponds to a kinetic energy of the condensed particles, the second term describes an interaction with an external potential corresponding to an optical or magnetic trap, and the last non-linear term corresponds to interatomic interactions treated within the framework of the mean-field and s-wave scattering approximations.  Gross-Pitaevskii equation gives a perfect description of the evolution of a conservative system with both the number of particles $N=\int |\psi(\mathbf{r},t)|^2d^2\mathbf{r}$ and the total energy:
\begin{align}
    H=\int \left[ \frac{\hbar^2}{2m}|\nabla\psi|^2+U|\psi|^2+\frac{g}{2}|\psi|^4\right]d^2\mathbf{r}.
\end{align}

The characteristic temperatures of the BEC phase transition for cold atomic gases lie in the nanokelvin temperature range, which is related to the large mass of condensing atoms, such as isotopes of sodium and rubidium. This motivated the search of analogues of BECs in condensed matter systems, where effective masses of various bosonic quasiparticles are several orders of magnitude smaller, and one can expect formation of quantum collective states at much higher temperatures.

One of the attractive possibilities is the use of ensembles of exciton polaritons (also known as cavity polaritons), which appear when a strong coupling regime between excitonic and photonic modes is realized in an optical microresonator \cite{Kavokin2017_OxfPr,Carusotto2013}. Being composite half-light half-matter particles, polaritons inherit an extremely small effective mass (about $10^{-5}$ of the mass of free electrons) and large coherence length (in the mm scale) \cite{Ballarini2017} from their photonic component.
However, the presence of an excitonic component results in efficient polariton-polariton interactions, which lead to a  strong nonlinear optical response. The polariton Bose-Einstein condensate and superfluidity have been experimentally observed at remarkably high temperatures under optical excitation \cite{Kasprzak2006,Balili2007,Amo2009,Lerario2017}. Polariton lasing, i.e., the generation of coherent polaritons in semiconductor cavities excited by electric current, has also been reported \cite{Schneider2013}.

In addition to the difference in effective masses, there are several important distinctions between polaritonic and atomic systems. 

First, polaritons can be directly created optically and have finite lifetimes because of the possibility for the photons to leave the system through the partially transparent Bragg mirrors. 

Second, the presence of the excitonic fraction in a polariton makes possible efficient polariton-phonon interaction, which can both couple polaritons to an ensemble of incoherent excitons and lead to the energy relaxation within the polariton liquid itself. 

These differences require substantial modifications of the dynamic equation \ref{GPEq}. The finite lifetime of polaritons and an external coherent pump can also be introduced straightforwardly as a simple linear decay term and a complex time exponent source on the right-hand side of the equation \ref{GPEq} \cite{CarusottoCiuti2004}. The incoherent pump and the coupling between polaritons and incoherent excitons are usually described in the framework of the Wouters-Carusotto model, which was first formulated for scalar polaritons \cite{WoutersCarusotto2007} and then generalized for the \mbox{spinor case \cite{Borgh2010}}. 

The presence of dissipation requires an external pump that creates new particles, and the steady state is determined by the balance between gain and loss. From a dynamical perspective, this implies the existence of attractors in the system. The properties of dissipative condensates can differ significantly from those of their conservative counterparts. For example, the dispersion of bogolons in a dissipative condensate is complex, with eigenenergies acquiring nonzero imaginary parts. This highlights why the dynamics of dissipative condensates represent a large and rapidly developing area of research.

Beyond gain and loss mechanisms, another crucial aspect of polariton condensate dynamics is energy relaxation, which plays an important role in the formation of steady states and the redistribution of momentum. It has been studied extensively, with models incorporating coupling to thermal reservoirs and energy-dependent gain \cite{wouters2010,wouters2010energy}. Another approach, following the original idea of Pitaevskii \cite{Pitaevskii1959}, incorporates dissipation phenomenologically by making the Hamiltonian non-Hermitian as \(\hat{H} \rightarrow (1 - \Lambda)\hat{H}\), where \(\Lambda\) is a small dimensionless parameter characterizing the dissipation strength \cite{Solnyshkov2014}. However, a closer inspection shows that such energy-dependent damping alters both the amplitude and the phase of the macroscopic wavefunction, leading to a non-conservation of the total number of particles. Therefore, this method cannot be interpreted as a pure energy relaxation mechanism, as it simultaneously induces particle loss. A consistent description of energy relaxation in a closed condensate system requires dissipation terms that reduce the energy while preserving the norm of the wavefunction.

The present paper represents an attempt to construct the theory of pure energy relaxation in a polariton system. Our analysis is based on the quantum hydrodynamics approach, describing the system dynamics via set of classical field Hamilton equations for canonically conjugated variables of concentration and phase two and introducing relaxation in a natural way by adding the gradient term. We analyze how this term affects the dynamics of the polariton droplets and dispersion of the elementary excitations. We consider the scalar case, focusing on pure energy relaxation only, leaving the spinor case and a description of polarization and spin relaxation for follow-up work.   

Our goal here is not to present a modeling of system dynamics in any particular experimental configuration, but rather to focus on the fundamental role of the damping mechanisms, which were overlooked before. Therefore, for the reason of clarity of the presentation, in our analysis we neglect all other dissipative processes, such as finite lifetimes, external pumping, and coupling with an incoherent excitonic reservoir, for which well established theoretical approaches exist already. These terms can be easily taken into account when analyzing a particular set of experimental data. 

\textit{Dynamic equations.} Let us note that the conservative Gross-Pitraevskii equation \ref{GPEq} is nothing but an equation for a classical field, which can be obtained using the least action $\delta S=0$, $S=\int \mathscr{L} d^2\mathbf{r}dt$ with the Lagrangian being (we take $U=0$ and the spatial dimensionality two characteristic for polariton systems):
\begin{widetext}
\begin{align}
    &\mathscr{L}=\frac{i\hbar}{2}\left(\psi^*\frac{\partial\psi}{\partial t}-\psi\frac{\partial\psi^*}{\partial t}\right)-\frac{\hbar^2}{2m}(\nabla\psi^*)(\nabla\psi)-\frac{g}{2}\left(\psi^*\psi\right)^2.
\end{align}
\end{widetext}
Using the Madelung representation of a field function in terms of density-phase variables,
\begin{align}
    &\psi=\sqrt{\rho}e^{-i\theta}, \label{Madelung}
\end{align}
one gets:
\begin{align}
    &\mathscr{L} = \hbar \rho \partial_t \theta - \frac{\hbar^2}{2m} \left[(\nabla \sqrt{\rho})^2 + \rho(\nabla \theta)^2\right] - \frac{g}{2}\rho^2.
\end{align}

From this expression it follows that the angle $\theta$ can be considered as a generalized field coordinate, while the density $\rho$ corresponds to canonically conjugated momentum as:
\begin{align}
   \pi=\frac{\partial\mathscr{L}}{\partial(\partial_t\theta)}=\hbar\rho
\end{align}
with the canonical field Hamiltonian being
\begin{align}
     &\mathscr{H}=\pi\partial_t\theta -\mathscr{L}= \frac{\hbar^2}{2m} \left[(\nabla \sqrt{\rho})^2 + \rho (\nabla \theta)^2\right] +  \frac{g}{2}\rho^2.
\end{align}

The dynamic field equation can be thus represented in the Hamiltionan form as:

\begin{align}
    &\partial_t \pi=\hbar \partial_t \rho=-\frac{\delta\mathscr{H}}{\delta\theta}=\frac{\hbar^2}{m} \nabla(\rho \nabla\theta),\label{EqRho}\\
     & \partial_t \theta=\frac{\delta\mathscr{H}}{\delta\pi}= \frac{\hbar (\nabla \theta)^2}{2m}-\frac{\hbar}{2m\sqrt \rho} \nabla^2 \sqrt{\rho} + \frac{g}{\hbar}\rho.\label{EqPhi}
\end{align}

This system, of course, is fully equivalent to Eq.\ref{GPEq} (with $U=0$) and can be obtained from it directly using the substitution \ref{Madelung}. However, it has a very important advantage as pure energy relaxation can be directly introduced to it, as it is discussed below.

Let us first note that Eq. \ref{EqRho} is nothing but a continuity equation for the conserving quantity $\rho$ with current density
\begin{equation}
\mathbf{j}=-\frac{\hbar\rho}{m}\nabla\theta.    
\end{equation}
 Pure energy relaxation should not affect particle number conservation, so we can do nothing but leave this equation as is.

Energy relaxation, which occurs due to various physical processes—such as the interaction of polaritons with phonons—can only be incorporated through additional terms in Eq.~\ref{EqPhi}. A natural way to phenomenologically introduce energy relaxation is by adding the term $-\delta\mathscr{H}/\delta \theta$ to Eq.~\ref{EqPhi}, which drives the system toward a state of minimal energy. The resulting modified dynamical equations take the form:
%
%
\begin{align}
    &\hbar \partial_t \rho=-\frac{\delta\mathscr{H}}{\delta\theta}=\frac{\hbar^2}{m} \nabla(\rho \nabla\theta),\label{EqRRho}\\
    &\partial_t  \label{EqTheta} \theta=\frac{\delta\mathscr{H}}{\delta\pi}-\gamma\frac{\delta\mathscr{H}}{\delta\theta}=\\ \nonumber 
    &=\frac{\hbar (\nabla \theta)^2}{2m}-\frac{\hbar}{2m\sqrt \rho} \nabla^2 \sqrt{\rho} + \frac{g}{\hbar}\rho +  \lambda \nabla(\rho \nabla\theta),  
\end{align}
where $\lambda=\hbar^2\gamma/m$ is a phenomenological energy damping constant. The energy of the system is not any more constant but decreases with time as
\begin{widetext}
\begin{align} 
    &\frac{dE}{dt}=\int \frac{\partial \mathscr{H}}{\partial t}d^2\mathbf{r}=-\frac{\lambda \hbar^2}{m}\int\left[\nabla(\rho \nabla\theta)\right]^2d^2\mathbf{r}=-\lambda m \int|\nabla\mathbf{j}|^2d^2\mathbf{r}\leq 0. \label{EqEnergy}
\end{align}
\end{widetext}
The system thus relaxes to the minimal energy state, conserving the total number of particles.

Note that we can rewrite the equations \ref{EqRRho},\ref{EqTheta} back to a single equation for the field function \ref{Madelung} as follows:
\begin{widetext}
\begin{align} \label{Finaleq}
    &i\hbar\frac{\partial\psi}{\partial t} =-\frac{\hbar^2}{2m}\nabla^2\psi+g|\psi|^2\psi-m\lambda\psi\nabla\mathbf{j}=-\frac{\hbar^2}{2m}\nabla^2\psi+g|\psi|^2\psi+\frac{i\hbar}{2}\lambda\psi\left[\psi^\ast\nabla^2\psi-\psi\nabla^2\psi^\ast\right]. 
\end{align}
\end{widetext}

This equation constitutes the main result of the current work. In the following, we show that the energy relaxation crucially affects both the dispersion of elementary excitations and the spatio-temporal dynamics of polarition droplets.

\textit{Dispersion of elementary excitations and superfluidity.} It is a known fact that the dispersion of elementary excitations of a conservative spatially homogeneous condensate with density $\rho_0$ is gapless and linear in $\mathbf{k}$, and at small momenta is described by the widely known Bogoliubov formula $\omega(k)=\sqrt{g\rho_0/m} k$, which according to the Landau criterion corresponds to the onset of superfluidity in the system, with 
\begin{equation}
    v_{c}=\sqrt{\frac{g\rho_0}{m}}   \label{critical_vel}
\end{equation}
being a critical velocity. 

This result can be obtained directly from Eq. \ref{GPEq} by calculating the dispersion of small excitation on the background of the spatially uniform condensate. We can apply a similar procedure to analyze how the pure energy relaxation will affect the dispersion.

\begin{figure}[t!]
\includegraphics[width=1.0\linewidth]{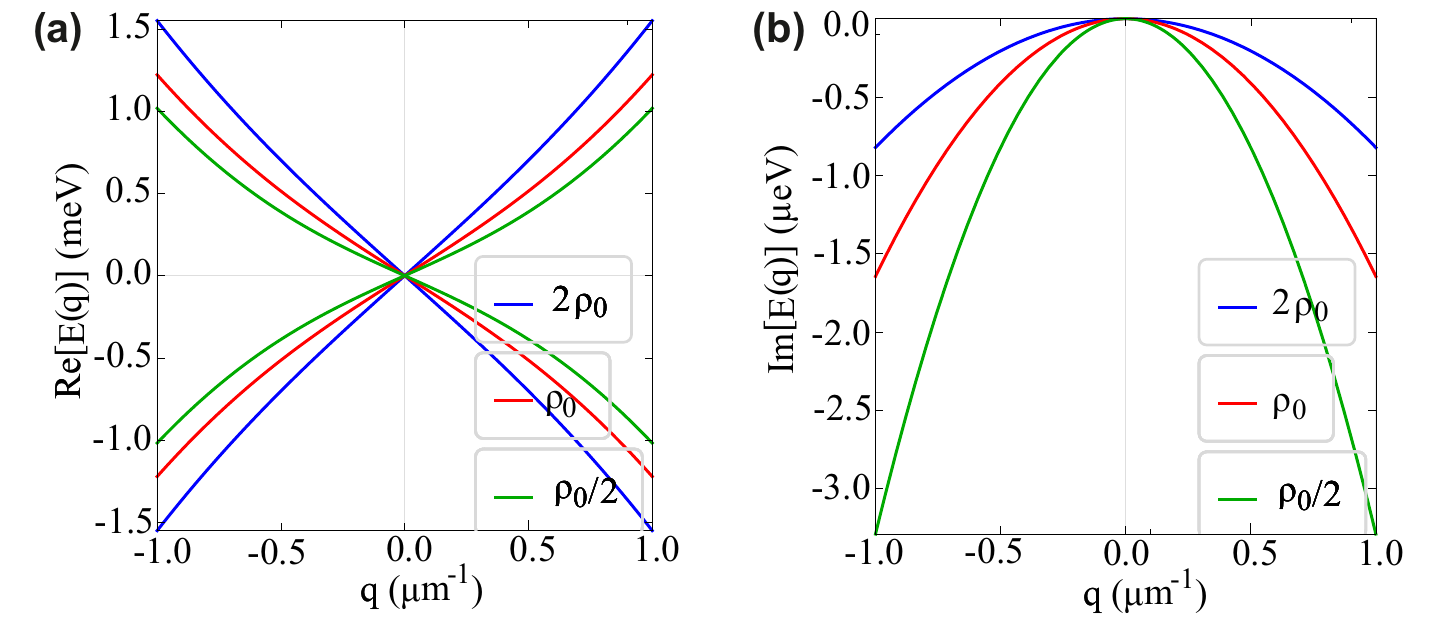}
\caption{The real (a) and imaginary (b) parts of the dispersion $E(q)=\hbar \omega(q)$ for different condensate densities $\rho_0$. The condensate is resting in the laboratory reference system (the phase gradient of $\psi$ is equal to zero). The real part of dispersion shows linear scaling with $q$ at small momenta, characteristic of superfluid behavior. The critical superfluid velocity increases with the density, as expected. The imaginary part of dispersion is responsible for the decay of the excitations provided by the pure energy relaxation term and scales quadratically with $q$. Parameters of the system are $g = 6 \cdot 10^{-3} \ meV \cdot \mu m^2 $, $m=5 \cdot 10^{-5}m_{0}$, $\rho_{0}= 100 \ \mu m^{-2}$, $\lambda=0.5 \cdot 10^{-4} \mu \mathrm{m^4}\cdot\mathrm{ps}^{-1}$.} \label{dispersion}
\end{figure}

The total field corresponding to a condensate of a density $\rho_0$ and an excitation with wavevector $\bf{q}$ and frequency $\omega$ is:
\begin{equation}
    \psi(\mathbf{r},t)=e^{-i \frac{g\rho_0 }{\hbar}t} \left[\sqrt{\rho_0}+ \xi e^{i(\mathbf{q}\cdot\mathbf{r}-\omega t)}+\bar \xi^\ast e^{-i(\mathbf{q}\cdot\mathbf{r}-\omega t)}\right], \label{AnsatzBogoliubov}
\end{equation}
where $\xi, $ and $\bar \xi$ are the amplitudes of the small perturbations, $|\xi|^2, \, |\bar \xi|^2 \ll \rho_0$. Placing ansatz Eq.\ref{AnsatzBogoliubov} in Eq.\ref{Finaleq} and linearizing it, one gets the following for the dispersion of elementary excitation:
\begin{equation}
\omega(q)=\sqrt{ \frac{g\rho_0}{m} q^2       +\left(\frac{\hbar^2}{4m^2}-\frac{\lambda^2 \rho_0^2}{4}\right)q^4} - i\frac{\lambda \rho_0 q^2}{2},     \label{EqDispersion}
\end{equation}
The real and imaginary parts of the dispersions are shown in Figure \ref{dispersion}. Note that in small $q$ the real part of the dispersion remains linear $\partial \omega/\partial q |_{q=0}  =v_c$ and therefore, according to the Landau criterion, the condensate is superfluid with critical velocity $v_c$ independently of $\lambda$. The negative imaginary part is responsible for the decay of the excitations provided by the pure energy relaxation term, which scales quadratically with $q$ and is proportional to the condensate density $\rho_0$, which, as expected, reflects the effect of bosonic stimulation.

Let us remark that both the effect of polariton superfluidity \cite{Amo2009,Lerario2017} and the formation of a linear polariton dispersion above the condensation threshold in a dissipative system was reported experimentally \cite{HoflingBogoliubov,Claude2023}. Note also that models of dissipative polariton fluids without pure energy relaxation give a qualitatively different dispersion, with flat regions in energy bands \cite{WoutersCarusotto2007,SolnyshkovDispersion}, and no clean superfluid behavior \cite{Amelio2020}.

The onset of a superfluid behavior in our model can be directly tested numerically. Consider an obstacle moving across a uniform condensate corresponding to a field function $\psi=\sqrt{\rho_0}$ and described by the external potential in the form of a Gaussian function $V=V_0 \mathrm{exp}\left[-(x-v_p t)^2/w_p^2\right]$ where $V_0$, $w_p$ and $v_p$ are the depth, width and velocity of the potential. 

\begin{figure}[t!]
\begin{center}
\includegraphics[width=1.0\linewidth]{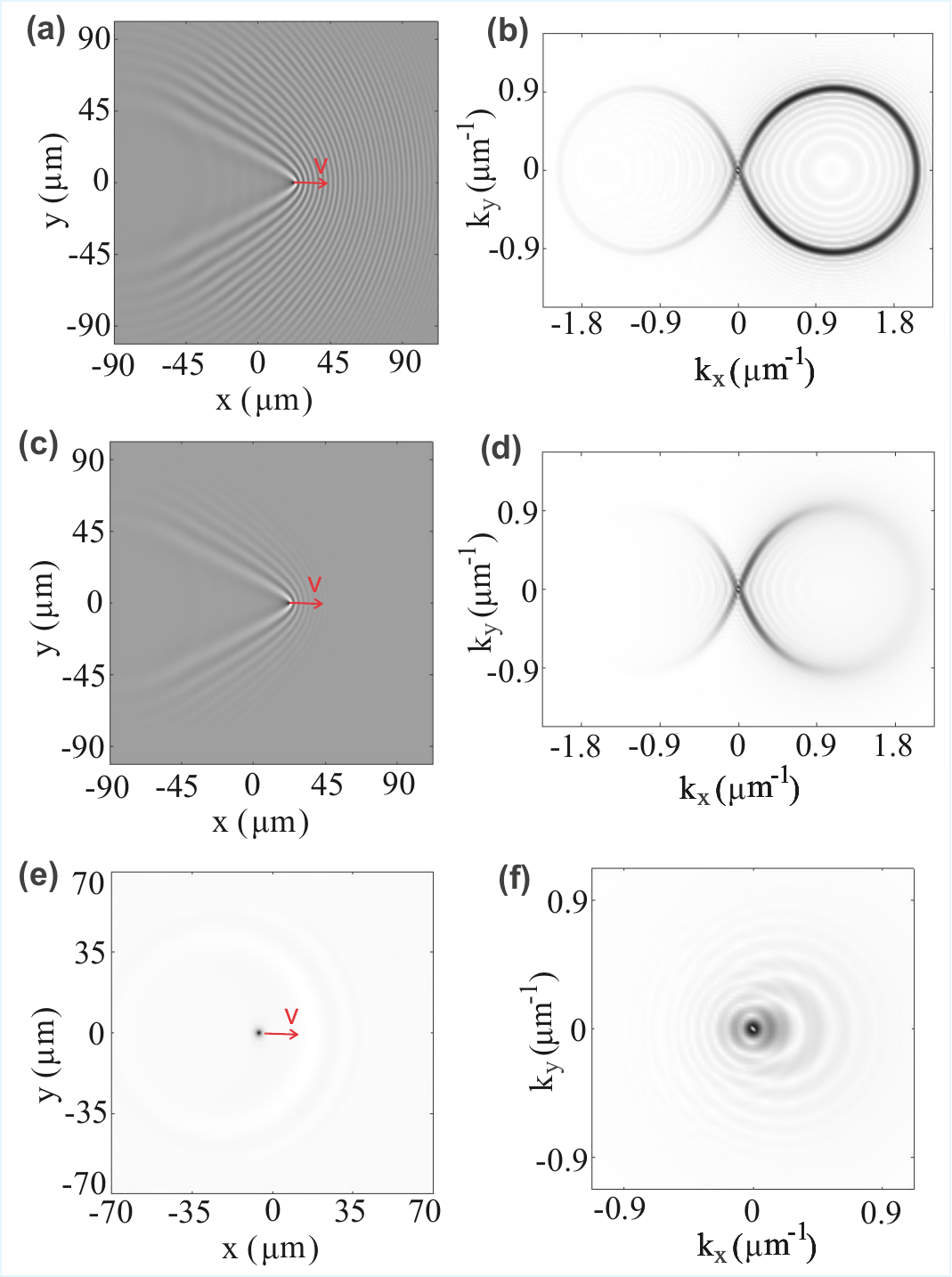}
\end{center}
\caption{The density (a) and the spatial spectrum (b) of a supersonic condensate in the presence of an obstacle moving across it, with perturbation potential being $V=V_0 \mathrm{exp}\left[-(x-v_p t)^2/w_p^2\right]$.  The density of the condensate is $\rho_0=50 \ \mu m^{-2}$, $V_0=0.3 \ meV$, $w_p=0.56 \ \mu m$, $v_p=2.5 \ \mu \mathrm{m}\cdot\mathrm{ps}^{-1}$. The nonlinear polariton-polariton interaction \mbox{$g = 6 \cdot 10^{-3} \ m\mathrm{eV}\cdot\mu \mathrm{m}^2$}, the relaxation rate $\lambda=0$. Panels (c) and (d) show the same but in the presence of the relaxation $\lambda=1.4 \cdot 10^{-3} \mu \mathrm{m}^4\cdot\mathrm{ps}^{-1}$. Panels (e) and (f) show the same as (c) and (d), but for the subsonic case with  $v_p=0.5 \ \mu \mathrm{m}\cdot\mathrm{ps}^{-1}$.  \label{supersonic_obstacle}}
\end{figure}

The results of the 2D simulations are shown in Fig.~\ref{supersonic_obstacle}. The four upper panels illustrate the supersonic cases of obstacle motion in the absence (a),(b) and in the presence $\lambda=1.4 \cdot 10^{-3} \ \mu \mathrm{m}^4  \cdot \mathrm{ps}^{-1}$ of the energy relaxation.  The cones of the emitted waves are clearly seen in panels (a),(c). In the corresponding spatial spectra, see panels (b), (d), there are characteristic patterns corresponding to the phase matching condition $\mathrm{Re} [\omega(\mathbf{q})]=v_p q_x $, where $\bf{q}$ is a wavevector of a scattered wave. The difference introduced by the energy relaxation is that in this case the emitted waves slowly decay so that the field becomes localized (c), and the amplitudes of the scattered waves with high momenta are suppressed (compare panels (b) and (d)), since the decay rate of the linear excitations with $\bf q$ is proportional to $\mathrm{Im}[\omega(\bf q)]\sim q^2$.
In the case of the subsonic regime, no emitted waves have been observed, and this regime is very similar to that known for unperturbed GPE (see panels (e) and (f)).

\textit{Stability of superfluid and non-superfluid flows.}

\begin{figure}[!t]
\begin{center}
\includegraphics[width=1.0\linewidth]{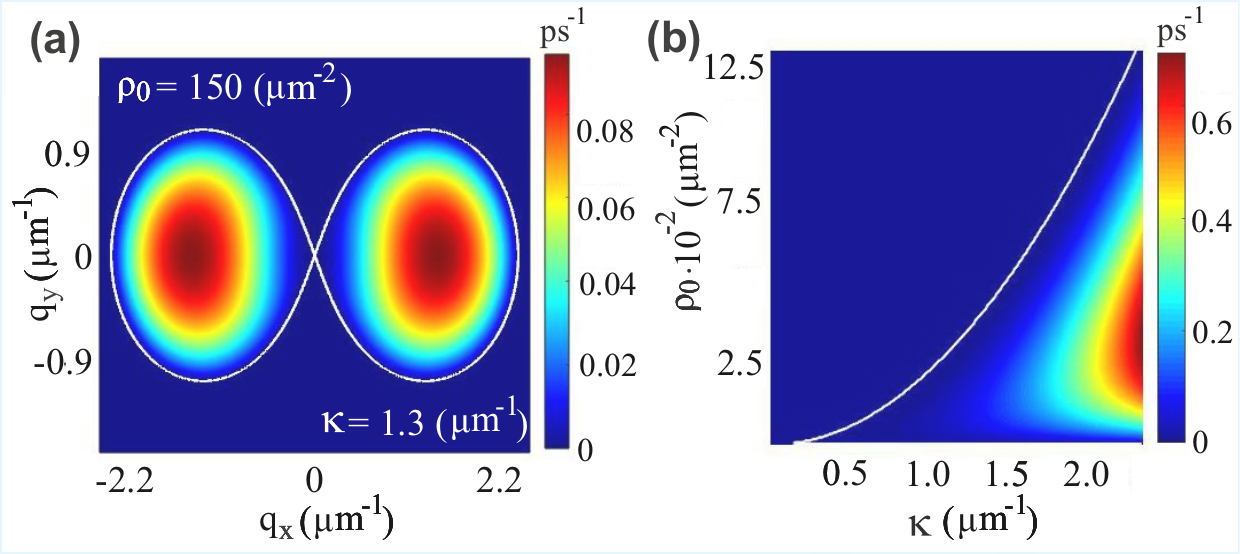}
\end{center}
\caption{Panel (a): increment rate of the perturbations characterized by the wavevectors $q_x$ and $q_y$ (imaginary part of $\omega(\vec q)$ given by (\ref{EqDispersion})). The density of the condensate is $\rho_0=150 \ \mu \mathrm{m}^{-2}$ , the condensate propagates along the $x$ axis with the wavevector $\kappa=1.3  \ \mu \mathrm{m}^{-1}$. The parameters are \mbox{$g=  6 \cdot 10^{-3} \ m\mathrm{eV}\cdot \mu \mathrm{m}^2 $}, $\lambda=1.4 \cdot 10^{-3} \mu \mathrm{m^4}\cdot\mathrm{ps}^{-1}$. The white lines show the boundary where the increments roll to zero. 
Panel (b): maximum increment of the instability as a function of the condensate density $\rho_0$ and its wavevector $\kappa$. The white line shows the border separating the stable condensate (above the curve) and unstable condensate (below the curve). At $\rho=0$ the condensate is neutrally stable. \label{increment_moving_condensate}}
\end{figure}

All stationary solutions of GPE are also solutions of (\ref{Finaleq}). This follows from the fact that for that solution $\nabla \bf{j}=0$ and the term accounting for the relaxation vanishes. However, relaxation does affect the stability of solutions. To analyze this, one should substitute into Eq. \ref{Finaleq} the ansatz corresponding to a spatially uniform condensate propagates along the $x$ axis with the wavevector of the absolute value $\kappa$,
\begin{widetext}
\begin{equation}
    \psi(\mathbf{r},t)=e^{-i (\frac{g\rho_0 }{\hbar}+\frac{\hbar \kappa^2}{2m})t+i \kappa x} \left[\sqrt{\rho_0}+ \xi e^{i(\mathbf{q}\cdot\mathbf{r}-\omega t)}+\bar \xi^\ast e^{-i(\mathbf{q}\cdot\mathbf{r}-\omega t)}\right] \label{AnsatzBogoliubov2}
\end{equation}
\end{widetext}
 and then perform the linearization procedure. This gives the following dispersion relation for the excitations:
\begin{widetext}
\begin{equation}
\omega(q)=\sqrt{ \frac{g\rho_0}{m} q^2       +\left(\frac{\hbar^2}{4m^2}-\frac{\lambda^2 \rho_0^2}{4}\right)q^4 -\frac{i \hbar q^2}{m}\lambda \rho_0 \kappa q_x} - i\frac{\lambda \rho_0 q^2}{2} +\frac{\hbar}{m} \kappa q_{x},     \label{EqDispersion2}
\end{equation}
\end{widetext}
Note that for $\kappa=0$ Eqs.\ref{EqDispersion} and \ref{EqDispersion2} are equivalent. However, differently from Eq.\ref{EqDispersion} the imaginary part of Eq. \ref{EqDispersion2} can become positive, resulting in the development of the flow instability. The typical dependence of the instability increment (positive imaginary part of \ref{EqDispersion}) for an unstable condensate is shown in  Fig.~\ref{increment_moving_condensate}(a) as a function of $q_x$, $q_y$.
The dependency of the maximum instability increment as a function of $\rho_0$ and $\kappa$ is shown in  Fig.~\ref{increment_moving_condensate}(b). As expected, it is higher for fast-moving condensates. However, the dependency on the condensate density $\rho_0$ at fixed $\kappa$ is not monotonous: it is always zero for $\rho_0=0$, then grows, reaches its maximum, and then at some density is rolled to zero so that the condensate stabilizes. It happens exactly at the point where the velocity of the condensate becomes equal to the critical velocity defined by the Landau criterion. More details on the development of the instability are given in Supplementary Materials \cite{supplemental}.    

\textit{Deceleration of polariton droplets.} Let us now analyze how energy relaxation affects the dynamics of polariton droplets. To keep the presentation short, we consider the simplest case of one-dimensional Gaussian wavepacket $\psi(t=0, x)=\sqrt{\rho_m} \exp ( -x^2/w_0^2 +i k_0 x)$ with the maximum density $\rho_m$ and wavevector $k_0$. 

\begin{figure}[!t]
\begin{center}
\includegraphics[width=1.0\linewidth]{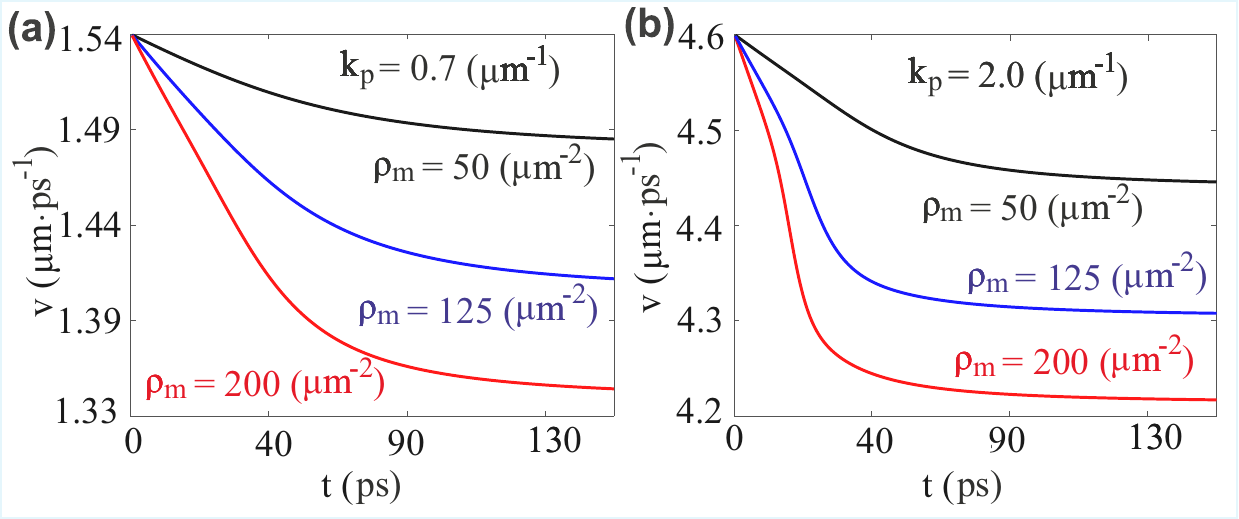}
\end{center}
\caption{Panel (a):  dependencies of the condensate droplet velocity $v$ on the propagation time $t$ for initial wavevector $k_0=0.7 \ \mu \mathrm{m}^{-1}$ and different initial peak densities $\rho_m=50 \ \mu \mathrm{m}^{-2}$ (black curve), $\rho_m= 125 \ \mu \mathrm{m}^{-2}$ (blue curve) and $\rho_m=200 \ \mu \mathrm{m}^{-2}$ (red curve). Panel (b) shows the same, but for the initial wavevector $k_0=2 \ \mu \mathrm{m}^{-1}$.  For all cases $g=0$ and $\lambda=1.4 \cdot 10^{-3} \mu \mathrm{m}^4\cdot\mathrm{ps}^{-1}$. \label{fig_velocity}}
\end{figure}

We have performed a series of numerical simulations that reveal that the non-zero relaxation energy rate $\lambda \neq 0$ results in the deceleration of the condensate pulse. The dependencies of the velocities of the condensates are defined as $v=d\langle x\rangle/dt$ with $\langle x\rangle=N^{-1} \int x |\psi|^2 dx$, $N=\int |\psi|^2 dx$
are shown in Fig.~\ref{fig_velocity} for different initial peak densities and velocities. 
As expected, the deceleration rate increases with the condensate densities; see Supplementary Materials \cite{supplemental}. We have checked that at short propagation times the velocity decreased exponentially, $v=v_0\mathrm{exp}(-\gamma t)$ with the decay rate $\gamma$ being independent on $k_0$ and scaling linearly with $\rho_m$. At longer propagation distances the pulse broadens, its intensity decreases, and the velocity scaling with time becomes polynomial, $v\sim t^{-2}$. Polartion-polariton interactions are repelling, so that they speed up the condensate spreading and thus suppress its deceleration. A detailed discussion of these effects will be presented elsewhere.

\textit{Conclusion} We have shown that the pure energy relaxation can be naturally introduced into the Gross-Pitaevskii equation. The resulting dissipation term conserves the number of particles and does not destroy the effect of superfluidity but strongly affects the dynamics of polariton droplets. Our results can also be applied to the cases of other bosonic condensates where the effects of energy relaxation are important.

We emphasize that the main objective of this paper is to demonstrate the importance of energy relaxation and to show that this effect can be accounted for by phenomenologically introducing a specific term into the Gross-Pitaevskii equation. The advantage of this approach is that the proposed term does not significantly complicate the mathematical model, while still capturing an effect of major physical relevance.

In this work, we considered the Gross-Pitaevskii equation with all dissipative terms omitted, except for the term representing energy relaxation. This was done intentionally to avoid masking the impact of energy relaxation by other effects, such as particle absorption and generation.

In order to accurately describe the dynamics of a real polariton condensate, multiple dissipative processes must be taken into account — including particle losses and condensation from a reservoir of incoherent excitons. One can expect that the interplay between these mechanisms and energy relaxation will strongly influence the condensate dynamics.

In this context, it is important to note that the derivation of the energy relaxation term presented in this work can be directly extended to the generalized Gross-Pitaevskii equation that incorporates a broader range of effects typical of dissipative polariton systems. We believe that including the proposed relaxation term in such generalized models will contribute to a more accurate theoretical description of experimentally observed phenomena in polariton condensates.

\bibliography{main}   

\end{document}


\preprint{APS/123-QED}

\title{Supplementary Materials: \\ Quantum hydrodynamics of a polariton fluid: pure energy relaxation terms}

\author{D. A. Saltykova}
\affiliation{ITMO University, St. Petersburg 197101, Russia}

\author{A. V. Yulin}
\affiliation{ITMO University, St. Petersburg 197101, Russia}

\author{I. A. Shelykh}

\affiliation{Science Institute, University of Iceland, Dunhagi 3, IS-107 Reykjavik, Iceland}

\date{\today}

\maketitle

\textit{ Resonant condition for the waves emitted in the supersonic regime} 

The dispersion relation (17) in the main text can be validated by comparing it with the results obtained from direct numerical simulations. We perform numerical simulations of the one-dimensional version of Eq. (14), perturbed by a moving potential given by
\begin{align}
    V=V_0 \exp(-\frac{(x-v_p t )^2}{w_p^2}),
\end{align}

where $V_0=0.3 \ meV$, $w_p=0.56 \ \mu m$. The initial conditions correspond to a stationary, spatially uniform condensate with a density of $\rho_0=50 \ \mu m^{-2}$. 

\begin{figure}[h!]
\begin{center}
\includegraphics[width=1.0\linewidth]{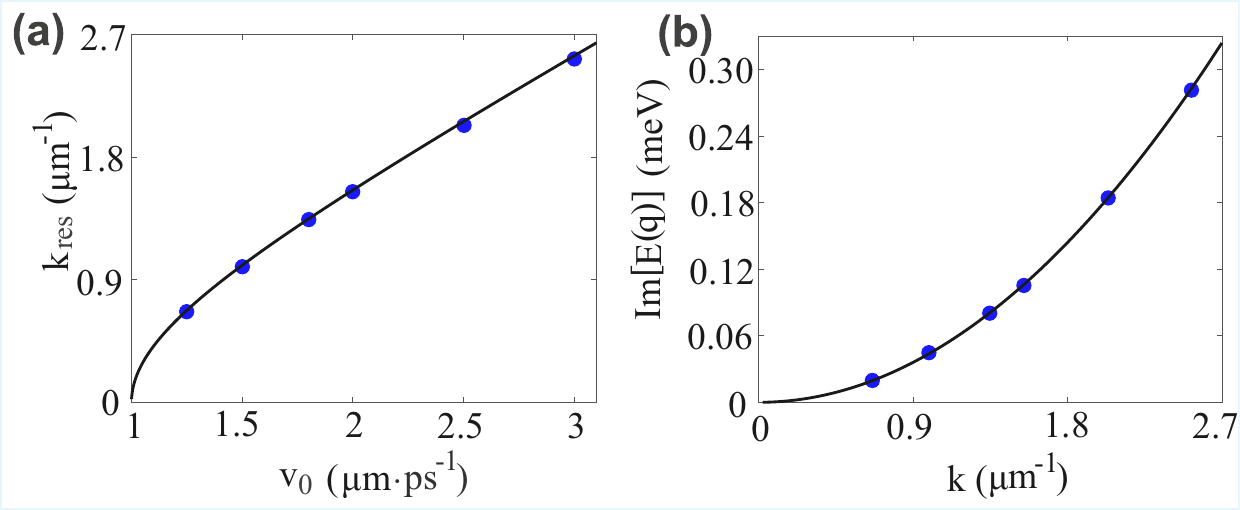}
\end{center}
\caption{ Panel (a) shows the dependence of the resonant wavevector of the emitted waves on the velocity of the moving localized potential obtained from the phase matching condition. The blue circles show the positions of the resonant spectral lines in the spatial spectrum of the condensate field obtained by direct numerical simulations of Eq.(14) in the main text. Panel (b) shows the dependency of the imaginary part of the dispersion characteristic (solid line) and the decay rate of the resonant radiation extracted from the numerical simulations.
The simulations and analytics are done for the spatially uniform condensate of density $\rho_0=50 \ \mu m^{-2}$, the polariton-polariton interaction $g =6 \cdot 10^{-3} meV \cdot \mu m^2$, the energy relaxation $\lambda=1.4 \cdot 10^{-3} \mu m^4 \cdot ps^{-1}$, the potential is given by  $V=V_0 exp(-\frac{(x-v_p t)^2}{w_p^2})$ with the depth $V_0=0.3 \ meV$ and width $w_p= 0.56 \ \mu m$.  \label{check_disper_suppl}}
\end{figure}

We observe that, when the velocity is sufficiently high, the potential generates a radiation tail, which, in the spectral domain, results in the emergence of a narrow resonant line. The position \(k_r\) of this spectral line can be determined from the dispersion characteristics by solving the phase-matching condition
\begin{align}
    Re [\omega(k_r)]=v_p k_r.
\end{align}
Thus, the resonant wavevector \(k_r(v_p)\) is a function of the velocity. Figure~\ref{check_disper_suppl}(a) shows this dependence as a solid curve, while the blue circles represent the positions of the resonant spectral lines derived from numerical simulations of Eq. (14). It is evident that all points align with the analytical curve, confirming that the numerical results are in excellent agreement with the theoretical predictions.
The decay rate of the resonantly emitted waves is examined by extracting the imaginary part of the dispersion characteristic from the numerical simulations of Eq. (14). We observe that in the stationary regime, the wave resonantly emitted by the moving obstacle decays exponentially in space, with a certain propagation length $L_{d}$ depending on the velocity of the obstacle. The relative velocity between the radiation and the potential is \( |v_g - v_p| \), where \( v_g = \text{Re}\left[\frac{\partial \omega}{\partial k} \bigg|_{k_r}\right] \) is the group velocity of the emitted waves. Thus, the spatial decay rate $\frac{1}{L_d}$ is related to the decay rate in time as
\begin{align}
   \frac{1}{L_d} = Im [\omega(k_r)] \frac{1}{|v_g-v_p|}.
\end{align}
 The imaginary part of the dispersion is extracted from the numerical simulations using the formula

\[
\text{Im}[\omega(k_r)] = \frac{|v_g - v_p|}{L_d}.
\]

Panel (b) of Fig.~\ref{check_disper_suppl} shows the eigenfrequencies calculated using this formula (solid line) and those obtained from the simulations (blue circles). The decay rate from the simulations is in good agreement with the analytical result.

\textit{The development of the instability}

We now examine the stability of the condensate moving at a velocity exceeding the critical velocity. When the velocity in the laboratory frame is below the critical value, the condensate remains stable, as predicted by \mbox{Eq.(19)} in the main text. This theoretical result agrees well with our numerical simulations of Eq.(14), where the initial conditions involve a spatially uniform condensate perturbed by weak noise. Over time, all excitations decay, and the system settles into a stationary state where the condensate moves at the same velocity as initially, within the accuracy of the simulations.

\begin{figure}[!t]
\begin{center}
\includegraphics[width=1.0\linewidth]{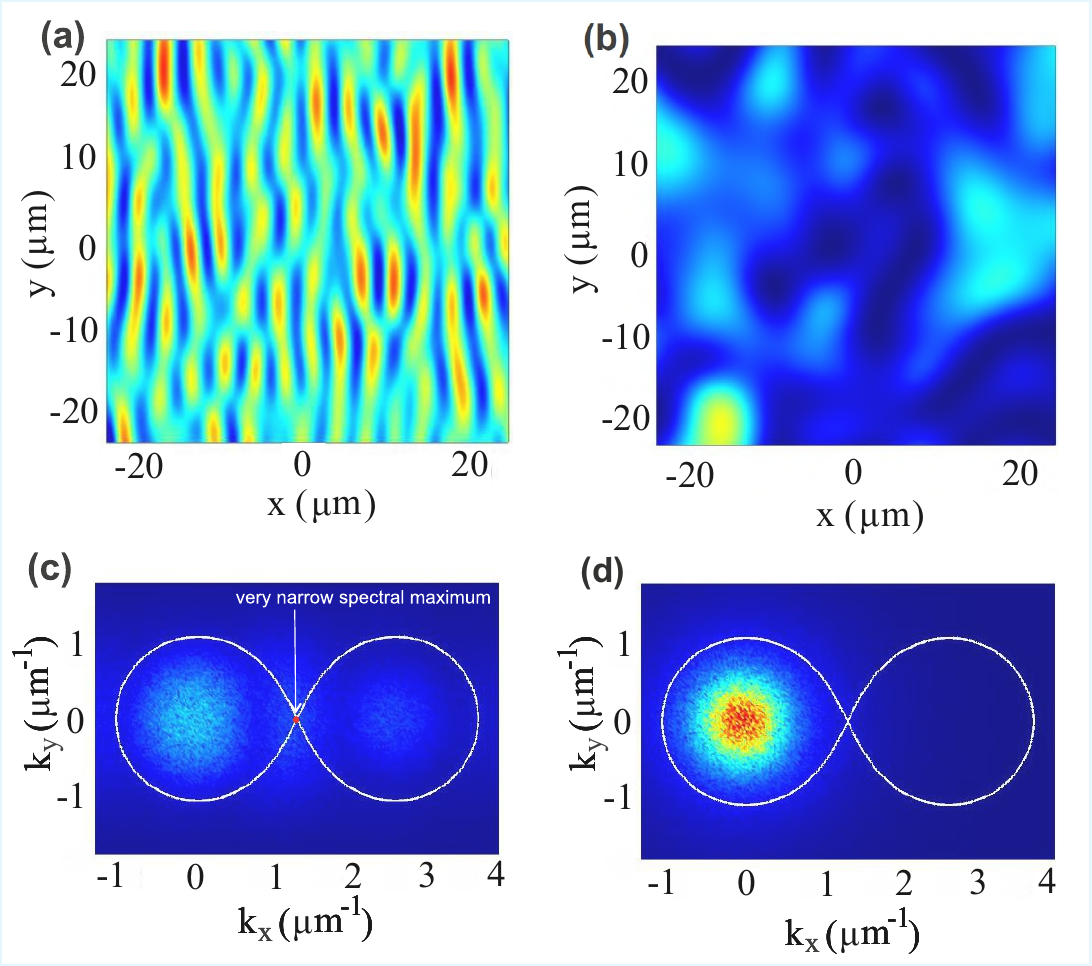}
\end{center}
\caption{The deviation of the condensate density from its mean value at \(t = 25 \, \text{ps}\) and \(t = 65 \, \text{ps}\). The initial conditions involve weak noise superimposed on a spatially uniform condensate with wavevector \(\kappa = 1.3 \, \mu\text{m}^{-1}\) along the \(x\)-axis. Panels (c) and (d) show the spatial spectra of the entire field at \(t = 25 \, \text{ps}\) and \(t = 65 \, \text{ps}\), respectively. The white curves in panels (c) and (d) mark the boundary of the instability region (inside the curve), calculated using Eq. (19). Other parameters: \(g = 6 \cdot 10^{-3} \, \text{meV} \cdot \mu\text{m}^2\), \(\lambda = 1.4 \cdot 10^{-3} \, \mu\text{m}^4 \cdot \text{ps}^{-1}\). \label{figure_2D_suppl}}
\end{figure}

We also observe that, in full agreement with Eq. (19), when the condensate moves at supersonic velocities, it becomes unstable and perturbations start to grow. This process is illustrated in Fig.~\ref{figure_2D_suppl}(a) and (b), which show the deviation of the condensate density from its average value at intermediate times (the quasi-linear stage of the instability) and at later times when the instability is nearly saturated.

Initially, vertically oriented stripes appear. The period of these stripes is defined by the wavevector at which the instability growth rate, given by Eq. (19), reaches its maximum. Panel (c) of the figure shows the spectrum of the field, which exhibits a very narrow peak at \mbox{\(k_x = 1.3 \, \mu\text{m}^{-1}\), \(k_y = 0\),} corresponding to the initial state of the condensate. The white curve indicates the boundary of the instability in \(k\)-space, and it is clear that spectral harmonics increase within this region.

At the fully developed stage of the instability, shown in panel (b), large-scale modulations of the condensate density remain, but the previously observed quasiperiodic stripe pattern has disappeared. The corresponding spectrum, shown in panel (d), is now strongly localized near \(k = 0\), indicating that the instability has led to a redistribution of momentum towards low-\(k\) modes, effectively decelerating the condensate.

\begin{figure}[!t]
\begin{center}
\includegraphics[width=1.0\linewidth]{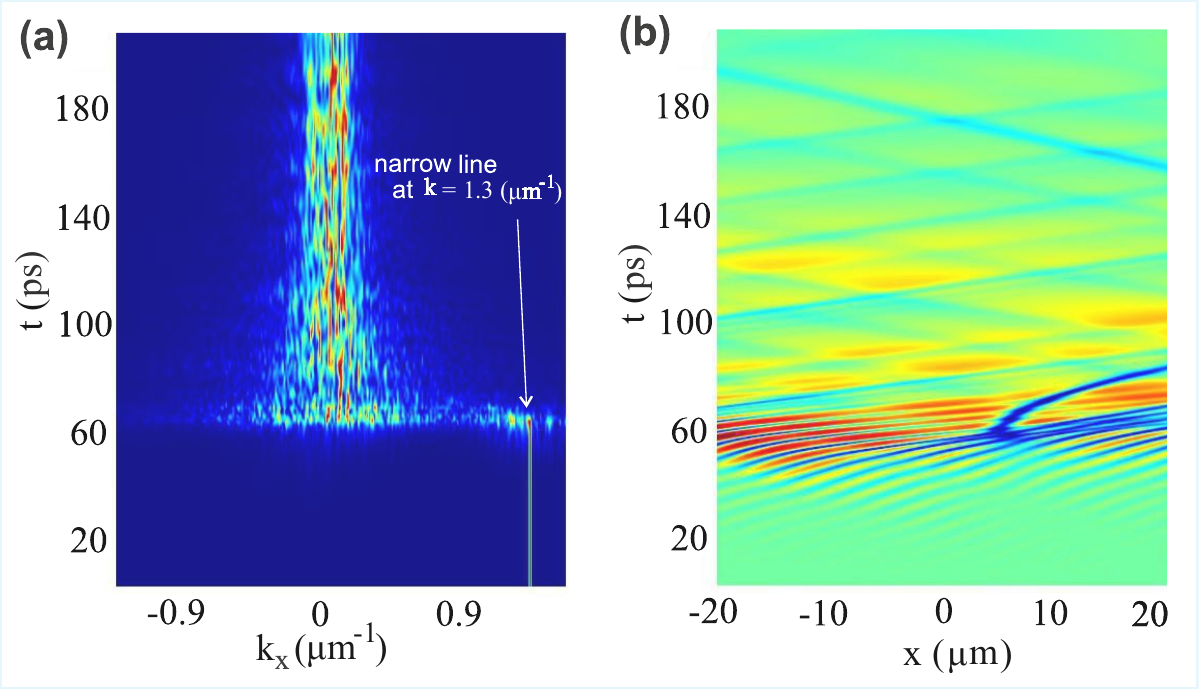}
\end{center}
\caption{The temporal development of the one-dimensional condensate with \(\kappa = 1.3 \, \mu\text{m}^{-1}\). The initial conditions correspond to a spatially uniform condensate with density \(\rho_0 = 50 \, \mu\text{m}^{-2}\), perturbed by weak noise. Panel (a) shows the evolution of the spectrum, while panel (b) shows the evolution of the field. Other parameters: \(g = 6 \cdot 10^{-3} \, \text{meV} \cdot \mu\text{m}^2\), \(\lambda = 1.4 \cdot 10^{-3} \, \mu\text{m}^4 \cdot \text{ps}^{-1}\). \label{figure_1D_instab_suppl}}
\end{figure}

The development of the instability in the one-dimensional condensate is shown in Fig.~\ref{figure_1D_instab_suppl}. At \(t \approx 65 \, \text{ps}\), quasiperiodic excitations begin to develop. In the spectral domain, this corresponds to the formation of a second, less intense, and much wider peak compared to the first. Eventually, the initial spectral line disappears, and only harmonics with small \(k\) survive. Notably, for \(g = 0\), short-scale excitations decay much faster than for larger \(g\), suggesting that some of these short-scale excitations may be dark solitons (the 1D analog of vortices) forming in the condensate.

\textit{The evolution of the field and the spectrum of the condensate droplets} 

\begin{figure}[!t]
\begin{center}
\includegraphics[width=1.0\linewidth]{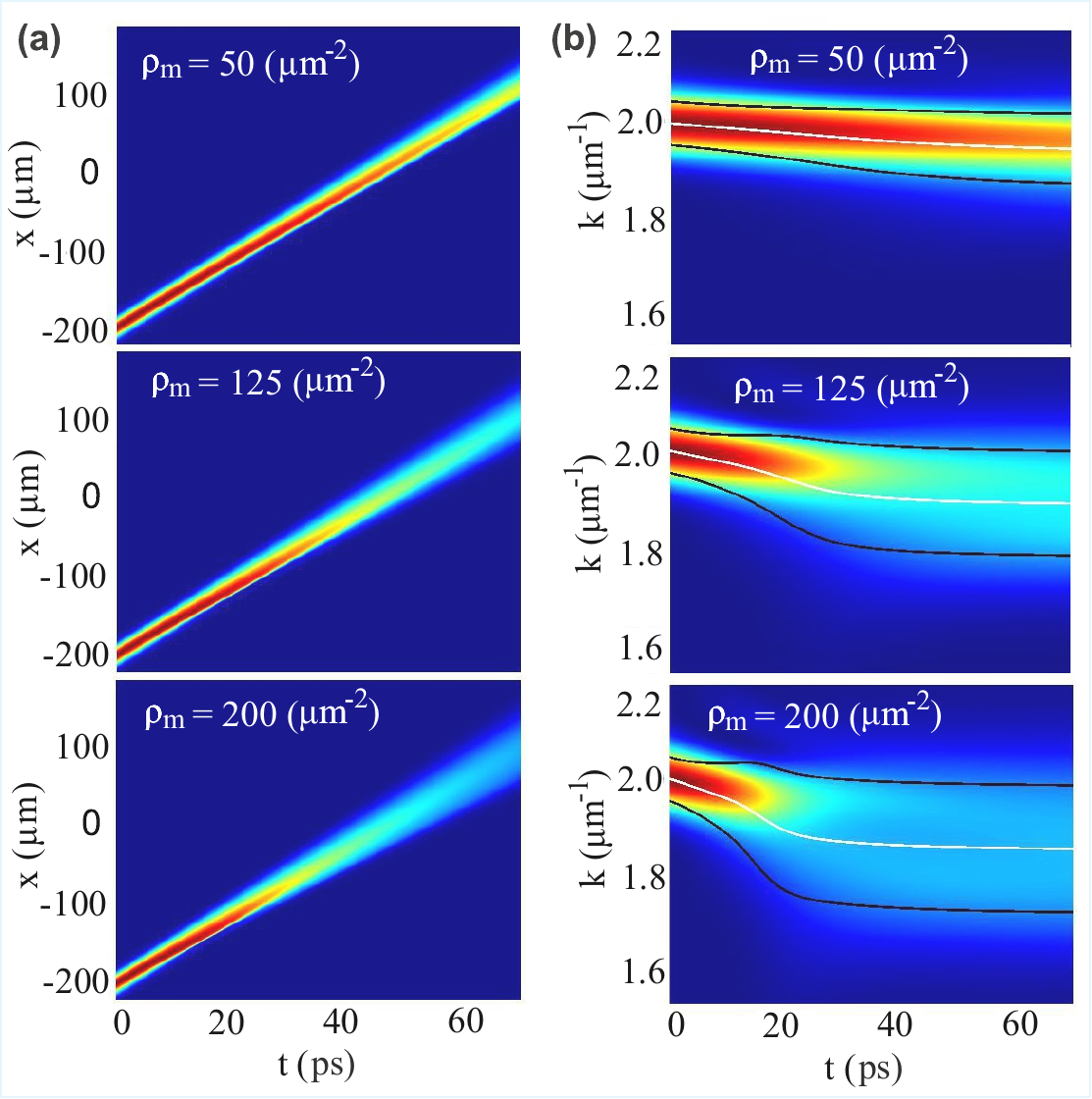}
\end{center}
\caption{Panel (a) shows the temporal evolution of the propagating droplet. At $t=0$ the field envelope is Gaussian $\psi(x, t=0)=\sqrt{\rho_m} \exp(-\frac{x^2}{w_p^2})\exp(i k_0 x) $, where $\rho_m$ is the initial maximum of the condensate density, $w_p=19.1 \ \mu m$ is the width of the droplet, and $k_0=2 \  \mu m^{-1}$ is its initial central wavevector.  The upper picture of the panel corresponds to $\rho_m=50 \ \mu m^{-2}$, the middle to $\rho_m=125 \ \mu m^{-2}$ and the bottom one to $\rho_m= 200 \ \mu m^{-2}$.  The evolution of the spatial spectra of the droplets is shown in panel (b). The white line marks the central wavevector defined as $k_c=\int k S(k) dk /N$, the black lines are $k_c \pm \Delta_{k \, S}$, where \mbox{$\Delta_{k \, S}=\sqrt{\int (k-k_c)^2 S(k) dk/N}$}. The physical meaning of the black lines is that they show the area filled by the condensate. The other parameters are $g=0$, $\lambda=1.4 \cdot 10^{-3} \mu m^4 \cdot ps^{-1}$. \label{figure_1D_evolution suppl}}
\end{figure}

Panels~(a) and (b) of Fig.~\ref{figure_1D_evolution suppl} show the evolution of the condensate field and its spectrum, respectively. As the condensate droplet propagates, its shape undergoes significant deformation, first becoming wider and then more asymmetric. Both the spreading rate and the development of asymmetry increase with the droplet's maximum density. In the spectral domain, we observe that the range of wavevectors occupied by the condensate shifts toward smaller \(k\) values during propagation. This indicates a deceleration of the droplet, as in these regimes, the droplet's velocity is closely approximated by the group velocity of linear waves calculated at the central wavevector.

\begin{figure}[!t]
\begin{center}
\includegraphics[width=1.0\linewidth]{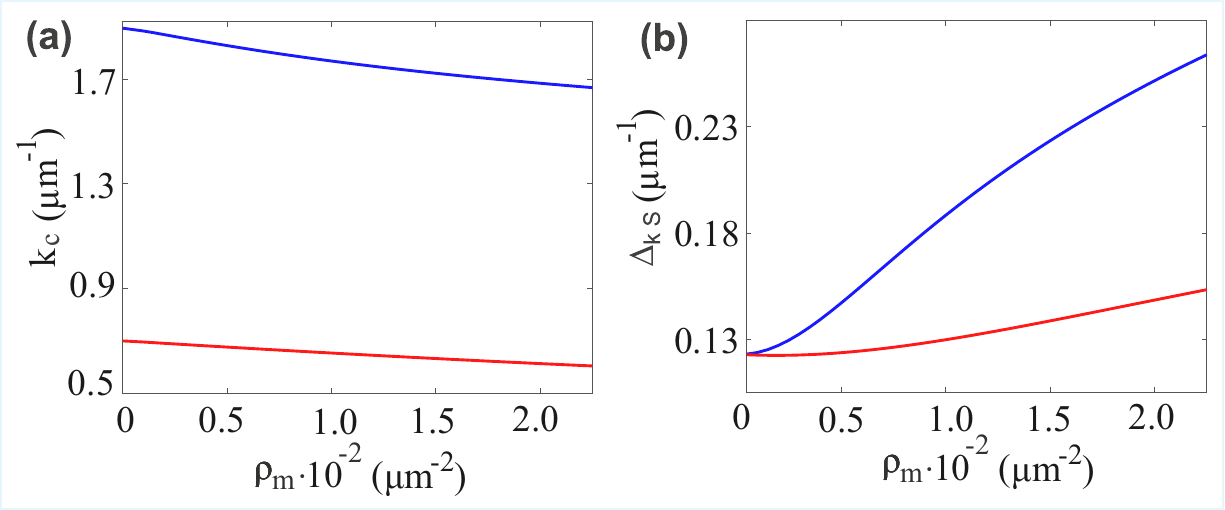}
\end{center}
\caption{Numerically calculated dependencies of the mean wavevector $k_c=\int k S(k) dk /N$ and the spectrum width $\Delta_{k \, S}=\sqrt{\int (k-k_c)^2 S(k) dk/N }$ on the maximum density $\rho_m$ of the initial condensate droplet are shown in panels (a) and (b) correspondingly. The parameters of the spatial spectrum of the condensate are measured after propagation time $t=75 \ ps$. The initial condensate field distribution is Gaussian $\psi(x, t=0)=\sqrt{\rho_m} \exp(-\frac{x^2}{w_p^2})\exp(i k_0 x) $ with the width $w_p=19.8 \ \mu m$.  The blue line corresponds to the droplets with the initial mean wavevector $k_0=0.7 \ \mu m^{-1}$, the red curves are for $k_0=2 \ \mu m^{-1}$.  The other parameters are $g=0 $, $\lambda=1.4 \cdot 10^{-3} \mu m^4 \cdot ps^{-1}$.
\label{K_and_DS_suppl}}
\end{figure}

To show how the density of the condensate affects the mean value of the wavevector and the width of the spatial spectrum of the condensate, we calculate the dependencies of these parameters on the maximum density of the initial distribution of the condensate after relatively long propagation time. The results are summarized in Fig.~\ref{K_and_DS_suppl}. Let us note that for the chosen parameters, the velocity of the condensate is equal to its wavevector with good accuracy. Our conclusion is therefore that energy relaxation, as expected, results in a decrease in the mean wavevector  and hence in a decrease in the droplet velocity. This effect also leads to a broadening of the spectrum. For energy relaxation to be a nonlinear effect, the spectrum modification is more pronounced for the droplets with higher maximum densities.

\begin{figure}[!t]
\begin{center}
\includegraphics[width=1.0\linewidth]{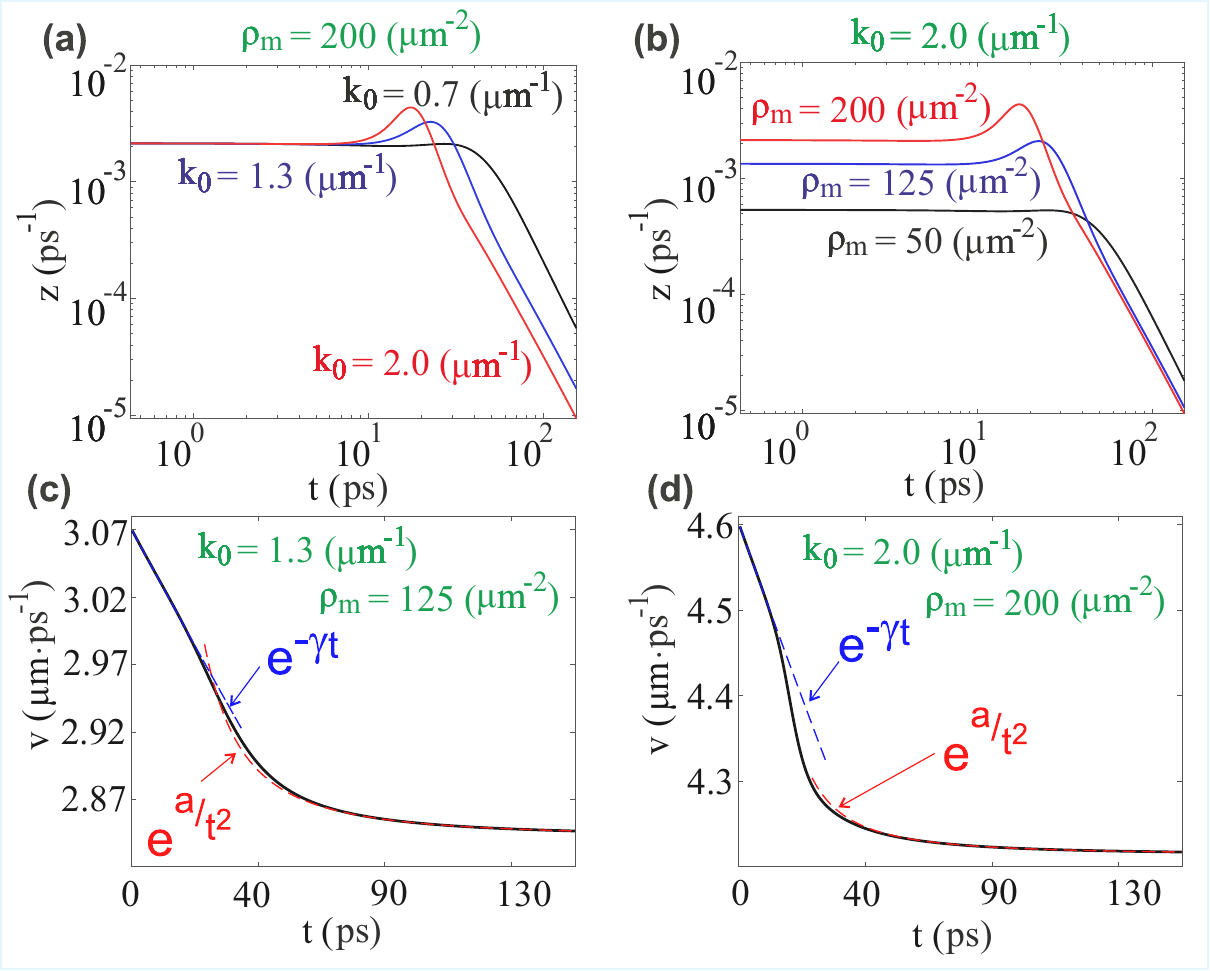}
\end{center}
\caption{The numerically calculated dependencies of the parameter $z=- \frac{\dot v}{v}$ plotted in double logarithmic scale are shown in panels (a) and (b). In panel (a), the maximum initial density is fixed to $\rho_m=200 \ \mu m^{-1} $ and the dependencies are plotted for the initial wavevectors of the condensate $k_0=2 \ \mu m^{-1}$ (red curve), $k=1.3 \ \mu m^{-1}$ (blue curve) and $k_0=0.7 \ \mu m^{-1}$ (black curve). Panels (b) show the dependencies $z(t)$ for fixed initial $k_0= 2 \ \mu m^{-1}$ and different maximum densities $\rho_m=200 \ \mu m^{-2}$ (red curve), $\rho_m=125 \ \mu m^{-2}$ (blue curve) and $\rho_m=50 \ \mu m^{-2}$ (black curve).  The dependencies of the velocities of the condensate droplets on time are shown in panels (c) and (d) for $k_0=1.3 \ \mu m^{-1}$, $\rho_m=125 \ \mu m^{-1}$ and $k_0=2 \ \mu m^{-1}$, $\rho_m= 200 \ \mu m^{-2}$ correspondingly. The fits $v \sim \exp(-\gamma  t)$ (blue curves) and $v \sim \exp(\frac{a}{t^2})$ (red curves) working at short and long propagation times are also shown in these panels. The other parameters are $w_p=19.1 \ \mu m$, $g=0$, $\lambda=1.4 \cdot 10^{-3} \mu m^4 \cdot ps^{-1}$.
\label{asympt_suppl}}
\end{figure}

Next, we study the evolution of the velocity of the Gaussian droplets and find that there are asymptotics at short and long times. To do this, we introduce the quantity $$z=- \frac{\dot v}{v}.$$
The dependencies of these parameters on time are shown in Fig.~\ref{asympt_suppl}(a),(b) on the double logarithmic scale for different initial wavevectors $k_0$ and maximum densities $\rho_m$ of the condensate. It is seen that in a short time the dependencies look as horizontal straight lines, and so $z$ can be approximated by a constant. This means that $\dot v = -\gamma v $ and, consequently, 
$$v(t)=v_0 \exp(-\gamma t).$$
Note that the equation governing the evolution of the droplet at short propagation times is equivalent to that describing a particle moving under the influence of viscous friction.

In our equations, the non-linear terms (energy relaxation) scale linearly with the condensate density, and from this we can expect that the deceleration rate $\gamma$ should also be proportional to $\rho_m$. We check that indeed this is the case, and the deceleration rate can be expressed as $\gamma=\gamma_0 \rho_m$. From our numerical simulations, we find that $\gamma_0 \approx 0.17 \ \mu m^2 \cdot ps^{-1} $ for the Gaussian pulse of the width $w_p=19.1 \ \mu m$.  

The fits, overlaid with the numerically calculated velocity-time dependencies of the droplet, appear in Fig.~\ref{asympt_suppl}(c) and (d). The fits accurately describe the velocity evolution at short propagation times. However, at longer times, dispersive spreading of the envelope becomes significant, and the fit is no longer applicable.

Surprisingly, there are asymptotic behaviors at long propagation times. As shown in Fig.~\ref{asympt_suppl}(a) and (b), the dependencies \(z(t)\) exhibit inclined straight lines on the double logarithmic scale at large \(t\). We found that

\[
z \sim \frac{1}{t^3}
\]

for large \(t\). Therefore, the velocity can be fitted as

\[
v = \tilde{v} \exp\left( \frac{a}{t^2} \right),
\]

where \(\tilde{v}\) is the limiting velocity and \(a\) is a constant determined from numerical simulations. The velocity decays to \(\tilde{v}\) as

\[
v = \tilde{v} + \frac{a \tilde{v}}{t^2}.
\]

By analyzing the fits overlaid with the numerically calculated \(v(t)\) dependencies (see Fig.~\ref{asympt_suppl}(c) and (d)), we observe that the fit accurately describes the velocity evolution.

\bibliography{main}